\documentclass[amsmath,amssymb,aps]{revtex4}
\usepackage{graphicx, epsfig}
\usepackage{latexsym}
\usepackage{epsfig}
\usepackage{latexsym}
\usepackage{amstext}
\usepackage{ulem}
\usepackage{amssymb}
\usepackage{array}
\usepackage{color}
\usepackage{graphicx}
\usepackage{dcolumn}
\usepackage{amsmath}
\usepackage{amssymb}

\newcommand{\ds}{\displaystyle}
\newcommand{\beq}{\begin{eqnarray}}
\newcommand{\eeq}{\end{eqnarray}}
\newcommand{\beqq}{\begin{eqnarray*}}
\newcommand{\eeqq}{\end{eqnarray*}}

\newcommand{\eps}{\varepsilon}

\newcommand{\n}{\mbox{\boldmath$n$}}

\def\ds#1{\displaystyle{#1}}

\begin{document}
\title{Asymptotics for the fastest among $N$ stochastic particles: role of an extended  initial distribution and an additional drift component}
\author{S. Toste$^1$, D. Holcman$^1$}
\affiliation{$^{1}$ Group of Data Modeling, Computational Biology, IBENS, Ecole Normale Superieure-PSL, France}
\begin{abstract}
We derive asymptotic formulas for the mean exit time $\bar{\tau}^{N}$ of the fastest among $N$ identical independently distributed Brownian particles to an absorbing boundary for various initial distributions (partially uniformly and exponentially distributed). Depending on the tail of the initial distribution, we report here a continuous algebraic decay law for $\bar{\tau}^{N}$, which differs from the classical Weibull or Gumbell results. We derive asymptotic formulas in dimension 1 and 2, for half-line and an interval that we compare with stochastic simulations. We also obtain formulas for an additive constant drift on the Brownian motion. Finally, we discuss some applications in cell biology where a molecular transduction pathway involves multiple steps and a long-tail initial distribution.
\end{abstract}
\maketitle

\section{Introduction}
\begin{figure}[http!]
\begin{center}
\includegraphics[scale = 0.4]{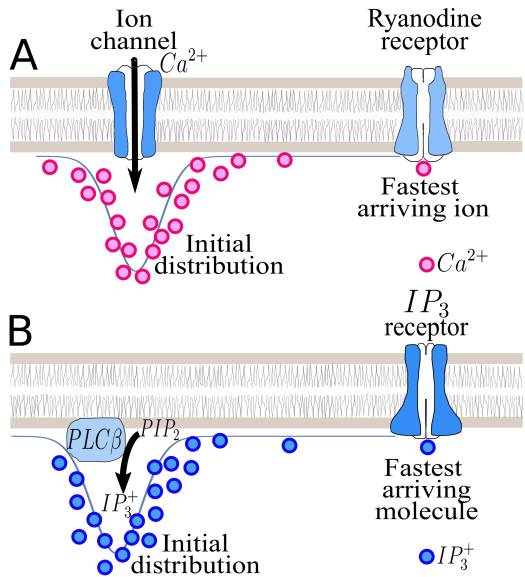}
\caption{\textbf{Two examples of fast molecular signalling where the initial distribution has a long tails.} \textbf{A.} Calcium ions enter very quickly through a channel or a cluster of channels. This fast entrance is associated to an initial distribution that can intersect at the tail with calcium sensitive receptors (such as Ryanodine receptors). \textbf{B.} IP$_3^+$ molecules, generated very quickly from PIP$_2$ at the cell membrane, that need to bind to IP$_3$-receptor.}
\label{graph0}
\end{center}
\end{figure}

Transient molecular activation in many cellular processes, such as gene transcription \cite{Purves:Book}, calcium activity in neuronal protrusion \cite{basnayake2018fast} or biochemical pathways associated with a secondary messenger transduction \cite{fain2019sensory} often occur in geometrical restricted micro-compartments, where the initial distribution of the source is well separated from the target site. To guarantee a reliable and fast activation, these processes are carried out by multiple redundant particles \cite{Holcmanschuss2018, coombs2019first,redner2019redundancy}. The multiplicity or redundancy has two effects: it increases the probability of finding a small target and, in parallel, decreases the mean activation time. Because it is usually costly to produce many copies of the same object, there is usually a compromise between the number of produced copies and the ultimate time scale of activation.  In addition, for molecular processes involving multiple time steps, as we shall see here, any possible spreading of the initial distribution can affect the final activation time. \\
{ For example, calcium ions enter in less than a few milliseconds inside a dendrite or neuronal synapses through few channels located on the membrane. After channels are closed, the calcium concentration has already spread, with a characteristic distribution, approximated as Gaussian in the diffusion limit (Fig. \ref{graph0}A). But other initial distributions are possible because cellular crowding could slow down diffusion, leading to anomalous diffusion \cite{metzler2000random} profiles. Starting with such long-tail instantaneous distribution, calcium ions can fulfil several functions such as activating buffer located at a certain distances away from the calcium channels. This step is necessary for the activation of a secondary messenger that can lead to change in the physiology. Indeed, it has been known for decades that calcium concentration is a key factor for the induction of long-term synaptic changes \cite{herring2016long}, but the exact reasons is still unclear: it could be to activate enough molecules quickly. We recently reported \cite{basnayake2018fast} that the initial distribution of injected calcium ions can modulate the probability of a calcium avalanche known as calcium-induced-calcium release, fundamental for the induction of physiological changes underlying learning and memory. Another generic example is the secondary biochemical messenger pathway (Fig. \ref{graph0}B): molecules such as cGMP, IP$_3^+$ or cAMP are generated near receptors and need to travel a certain distance away to activate a second pathway. In all these cases, the initial concentration of these molecules is critical for the genesis of rhythmic oscillations or the  amplification of a single molecular events.  As we shall see, the initial number and their distribution can be critical for the determination of the activation time of a transduction cascade.} \\
We recall that changing the initial number on the search time by the fastest has been quantified as follows: when there are $N$ i.i.d. particles,  generated at a specific point location (entrance of a channel or a receptor),  then the Mean Fastest Arrival Time (MFAT), which is the mean time it takes for the first one to arrive decays with $1/\log N$ (a classical property of the Gumbell distribution) \cite{weiss1983order,schehr2014exact,basnayake2018fast,majumdar2020extreme}, but the decays can be much faster $1/N^2$ when the initial distribution of particles is uniformly distributed \cite{redner2001guide,basnayake2019fastest,grebenkov2020single}.\\ Computing how the first arrival time depends on the initial numbers $N$ is key to formulate biophysical laws of activation by the fastest diffusing particles to reach a target. The type of motion could matter, as revealed for spermatozoa to arrive to the ovule location, modeled as persistent motion switching direction after hitting the surface \cite{yang2016search}. In general we are still missing the law of arrival for anomalous diffusion and many classical random motion such as for the full Langevin.\\
We compute here the mean arrival time $\bar{\tau}^{N}$ for the fastest among $N$ identical Brownian particles using short-time asymptotic of the diffusion equation. In particular, we consider the case of extended initial conditions. Indeed, as mentioned above, after a molecule is generated at a specific location, additional chemical processes are involved to provide the molecular activity required to interact with a given target. Specific reactions are phosphorylations in case of transcription factor or methylations \cite{alberts2015essential}. Indeed in some cases, transcription factors need to be phosphorylated or other molecules are needed, creating re-modeler complexes \cite{riedl2001phosphorylation}. During these specific activation, the initial molecule can move by small drift or diffuse away allowing the initial concentration to spread a bit, a situation that we are interested in here. \\
The manuscript is organized as follow. First, we recall the framework and derive explicit expressions for the MFAT when the initial distribution is uniform, intersecting or not with the target site for half a line $\mathbf{R}_{+}$ and for a segment $\Omega = [0,a]$. We study initial distributions of the form $p_1(x) = \frac{2b^{\frac{\alpha+1}{2}}}{\Gamma \left(\frac{\alpha +1}{2}\right)}x^{\alpha} e^{ -bx^2}$ with $b>0$ and $\alpha \geq 0$ and obtain the general formula (relation (\ref{genaral_alpha_formula}))
\beq
\bar{\tau}^{N}\sim  \frac{C_{\alpha}}{N^{\frac{2}{\alpha +1}}}
 \frac{1}{4Db} \text{   for $N\gg 1$},
\eeq
where $\Gamma$ is the Gamma function and
\beq
C_{\alpha}=\Gamma \left(\frac{\alpha +3}{\alpha +1}\right) \left(\frac{\sqrt{\pi}(\alpha +1)\Gamma \left(\frac{\alpha +1}{2}\right)}{2\Gamma \left(\frac{\alpha +2}{2}\right)} \right)^{\frac{2}{\alpha +1}}.
\eeq
This formula reveals a large spectrum of possible decay that depends on the analytical expression of the local overlap between the initial distribution and the target location. In addition, we provide an equivalent formula in two dimensions. Finally, we study the influence of a constant drift on the escape time $\bar{\tau}^{N}$.
\section{General framework for the first arrival time}
The shortest arrival time for $N$ non-interacting i.i.d. Brownian trajectories (molecules, proteins, ions) moving in a domain $\Omega$ to a small target (binding site) is defined as
\beq
\tau^1 = \min(t_1,..., t_N), \nonumber
\eeq
where $t_i$ are the i.i.d. arrival times of the $N$ particles. The complementary cumulative density function of $\tau^1$ is given by
\beq
Pr \left\lbrace \tau^1 >t \right\rbrace  = \left[Pr \left\lbrace t_1 >t \right\rbrace \right]^N, \nonumber
\eeq
where Pr$\left\lbrace t_1 >t \right\rbrace$ is the survival probability of a single particle prior to reaching the target. This probability can be computed from solving the diffusion equation
\beq
	\frac{\partial p(x,t)}{\partial t} &=& D \Delta p(x,t)\,\hbox{ for $x \in \Omega$, $t>0$} \nonumber \\
	p(x,0) &=& p_0(x) \,\,\, \text{for $x \in \Omega$} \nonumber \\
	\frac{\partial p(x,t)}{\partial \n}& = & 0 \, \hbox{ for } x \in \partial \Omega_r \nonumber \\
	p(x,t) &=& 0 \,\,\,\hbox{ for } x \in \partial \Omega_a \nonumber
\eeq
where $D$ is the diffusion coefficient and, the boundary $\partial \Omega$ contains $R$ binding sites $\partial \Omega_i \subset \partial \Omega$ with $i =1,..R$, we have then
\beq
\partial \Omega _a = \bigcup_{i = 1}^{R} \partial \Omega _i \nonumber
\eeq
and $\partial \Omega _r = \partial \Omega - \partial \Omega _a$. The survival probability is
\beq \label{survival1}
Pr\left\lbrace t_1>t\right\rbrace &=& \int_{\Omega} p(x,t) dx
\eeq
so that the probability density function (pdf) for the arrival of the first particle is
\beq\label{time dist}
	Pr \left\lbrace \tau^1 = t \right\rbrace  = \frac{d}{dt} Pr \left\lbrace \tau^1 < t \right\rbrace = N (Pr \left\lbrace t_1>t \right\rbrace)^{N-1} Pr\left\lbrace t_1 = t \right\rbrace,
\eeq
where the instantaneous probability is given by the probability flux
\beq
Pr \left\lbrace t_1 = t \right \rbrace  = \oint_{\partial\Omega_a} \frac{\partial p(x,t)}{\partial \textbf{n}}dS_{\textbf{x}}.\nonumber
\eeq
The Mean Fastest Arrival Time (MFAT) is defined as the mean time for the first among $N$ i.i.d. Brownian paths to reach the target and is obtained by computing the integral
\beq \label{mfpt}
	\overline{\tau}^N = \int_{0}^{\infty} Pr \left\lbrace \tau^1 > t \right\rbrace dt = \int_{0}^{\infty} \left[Pr \left\lbrace t_1 >t \right\rbrace \right]^N dt.
\eeq
\section{Arrival times for multiple initial distributions in dim 1}
\subsection{Arrival from a ray for multiple initial distributions}
We start with the case of a ray $\Omega = \mathbf{R}_{+}$, for which the solution of the diffusion equation
\beq \label{dim1condeltadirac}
\frac{\partial p(x,t)}{\partial t} &=& D \frac{\partial^2 p(x,t)}{\partial x^2} \text{for $x>0$, $t>0$}  \\
p(x,0) &=&\delta\left(x-y\right)  \hbox{ for $x >0$} \nonumber \\
p(0,t) &=& 0\hbox{ for $t>0$} \nonumber
\eeq
is given by
{\small \beq \label{p}
p(x,t) = \frac{1}{\sqrt{4 D \pi t}}\left[\exp\left\lbrace -\frac{(x-y)^2}{4Dt}\right\rbrace - \exp\left\lbrace -\frac{(x+y)^2}{4Dt}\right\rbrace\right].
\eeq
}
For a general initial condition, the solution of (\ref{dim1condeltadirac}) is the convolution of the initial pdf $p(x,0)$ with the elementary function presented in (\ref{p}). We previously treated the case of the Dirac delta function in dimension 1, 2 and 3 \cite{basnayake2019fastest,Basnayake2018,basnayake2018extreme} and also for initial distributions spread over a perpendicular segment in a cusp \cite{basnayake2020extremecusp}.\\
When the initial distribution of particles is uniform in a small portion of the ray, $[0;y_0]$, $p(x,0) = \frac{1}{y_0} \mathbb{I}_{\left[x \in [0,y_0] \right]}$,  then the solution is
{\small \beq
p(x,t)= \int_{0}^{y_0} \frac{1}{y_0 \sqrt{4Dt \pi}} \left[\exp\left( -\frac{(x-y)^2}{4Dt}\right) - \exp\left( - \frac{(x+y)^2}{4Dt}\right) \right] dy. \nonumber
\eeq
}
The survival probability given by relation (\ref{survival1}) is
\beq
Pr\{t_1>t\} = \int_{0}^{\infty} p(x,t) dx=
1 - \frac{2}{\pi} \int_{\frac{y_0}{\sqrt{4Dt}}}^{\infty} e^{-u^2}du + \frac{\sqrt{4Dt}}{y_0 \sqrt{\pi}}\left[e^{-\frac{y_0^2}{4Dt}}-1\right]. \nonumber
\eeq
For small time asymptotic $t\ll1$, $Pr\{t_1>t\} \sim  1-\frac{\sqrt{4Dt}}{y_0 \sqrt{\pi}}$ and thus using relation (\ref{mfpt}),
{\small
\beq \label{mfat unif [0,y_0] no drif}
\overline{\tau}^{N}
 \sim  \int_{0}^{\infty}\exp\left\lbrace -N\frac{\sqrt{4Dt}}{y_0 \sqrt{\pi}}\right\rbrace dt = \frac{y_0^2 \pi}{2D N^2}.
\eeq
}
This result shows that as soon as the initial condition overlaps with the target, the MFAT decay with order $1/N^2$. We now consider a local initially uniform distribution in the shifted interval $[y_1;y_2]$ where $y_1>0$, not overlapping with the target site. The initial normalized distribution function is thus $p_0(x) = \frac{1}{y_2 -y_1} \mathbb{I}_{\left\lbrace x \in [y_1,y_2] \right\rbrace}$ with $y_2>y_1>0$. We now compute the pdf for particle to reach the boundary. It is given by
\beq
	Pr\left\lbrace t_1>t\right\rbrace &=& \int_{0}^{\infty} \int_{y_1}^{y_2} \frac{1}{(y_2 -y_1) \sqrt{4Dt \pi}} \left[\exp\left\lbrace - \frac{(x-y)^2}{4Dt}\right\rbrace - \exp\left\lbrace - \frac{(x+y)^2}{4Dt}\right\rbrace \right] dy\,\, dx \nonumber \\
	& = & 1 - \frac{y_2}{y_2 -y_1}\left(\frac{2}{\pi} \int_{\frac{y_2}{\sqrt{4Dt}}}^{\infty} e^{-u^2}du\right) + \frac{y_1}{y_2 -y_1}\left(\frac{2}{\pi} \int_{\frac{y_1}{\sqrt{4Dt}}}^{\infty} e^{-u^2}du\right) \nonumber\\
	&+&\frac{\sqrt{4Dt}}{(y_2 -y_1) \sqrt{\pi}}\left[e^{-\frac{y_2^2}{4Dt}}-e^{-\frac{y_1^2}{4Dt}}\right]. \nonumber
\eeq
Expanding the complementary error function, we obtain for small $t$ asymptotic the relation
\beq \label{y1y2}
S(t)=Pr\left\lbrace t_1>t\right\rbrace	 \sim 1 - \frac{(\sqrt{4Dt})^3}{2(y_2 -y_1) \sqrt{\pi}}\left[\frac{e^{-\frac{y_1^2}{4Dt}}}{y_1^2}-\frac{e^{-\frac{y_2^2}{4Dt}}}{y_2^2}\right].
\eeq
{Note that expression (\ref{y1y2}) contains two exponentially small terms.  It is however possible to recover the case of an initial Dirac function by making the expansion $y_2 = y_1(1+\eps)$ and stuyding the limit when $\eps$ goes to zero in equation (\ref{y1y2}). In that case,  we have
\beq
S_{\eps}(t) \sim  1  - \frac{(\sqrt{4Dt})^3}{2y_1 \eps \sqrt{\pi}}\frac{e^{-\frac{y_1^2}{4Dt}}}{y_1^2} \left[1-\frac{e^{-\frac{y_1^2\left(2\eps + \eps^ 2\right)}{4Dt}}}{(1+\eps)^ 2}\right]. \nonumber
\eeq
A Taylor expansion in  $\eps$ and $\frac{\eps}{t}$ leads to
\beq
S_{\eps}(t) = 1 - \frac{(\sqrt{4Dt})^3 e^{-\frac{y_1^2}{4Dt}}}{\sqrt{\pi}y_1^3}\left[1+\frac{y_1^2}{4Dt} - \eps \left(1 + \frac{3y_1^2}{2\cdot 4Dt} +\frac{2y_1^4}{(4Dt)^ 2} \right) + O(\eps^2) +O\left(\frac{\eps}{t}\right) \right]. \nonumber
\eeq
When $\eps \rightarrow 0$, the survival probability $S_{\eps}(t)$ converges to $S_{0}(t)$ corresponding to an initial condition for the Dirac delta function at position $y_1$. However, the convergence is not uniform in $t$ in the interval $[0,\infty[$, preventing to use this expansion to estimate the MFAT for the case of an interval. Thus to leading order, using that
\beq
\underset{y_2 \rightarrow y_1}{\lim} Pr\left\lbrace t_1>t\right\rbrace	 \sim 1 - \frac{\sqrt{4Dt}}{\sqrt{\pi}}\left[\frac{e^{-\frac{y_1^2}{4Dt}}}{y_1}\right], \nonumber
\eeq
we obtain to leading order the asymptotic formula for $N\gg 1$
\beq \label{mfat unif [y_1,y_2] no drif}
\overline{\tau}_{\eps}^N \approx \frac{y_1^{2}}{4D \,\,\log\left(\frac{N}{\sqrt{\pi}}\right)+A_{\eps}},
\eeq
where $A_{\eps}=A_0+\eps A_1+..$, where $A_k$  are constants. Here $y_1$ is the shortest distance to the absorbing boundary. To conclude, to leading order, the MFAT for a small interval is the same as a Dirac delta function at the minimum point of the interval where the particles are uniformly distributed.}
\subsection{MFAT for an initial distribution asymptotically touching the target site} \label{section for the general distribution}
For an initial normalized distribution $p_1(x) = Kx^{\alpha} e^{ -bx^2}$ with $\alpha \geq 0$ and $b>0$, we have
$ K = \frac{2b^{\frac{\alpha+1}{2}}}{\Gamma \left(\frac{\alpha +1}{2}\right)}.$ The survival probability of the diffusion process is given by relation  (\ref{survival1}) leading to
\beq
Pr\left\lbrace t_1>t\right\rbrace &=& \frac{2b^{\frac{\alpha+1}{2}}}{\Gamma \left(\frac{\alpha +1}{2}\right)} \int_{0}^{\infty} \mathrm{erf} \left(\frac{y}{ \sqrt{4Dt}}\right) y^{\alpha} \cdot \exp\left\lbrace -by^2\right\rbrace dy \nonumber \\
&=&\frac{2}{\Gamma \left(\frac{\alpha +1}{2}\right) \cdot \left(4Dbt\right)^{\frac{1}{2}}} \frac{\Gamma \left(\frac{\alpha +2}{2}\right)}{\sqrt{\pi}} \,_2F_1\left[\frac{1}{2},\frac{\alpha +2}{2}, \frac{3}{2}, -\frac{1}{4Dbt} \right]\nonumber
\eeq
where $\,_2F_1\left[a,b,c,z \right]$ is the Gauss hyper-geometric function. The expansion for $z \rightarrow \infty$ ($t$ small), gives
\beq
\,_2F_1\left[\frac{1}{2},\frac{\alpha +2}{2}, \frac{3}{2}, -\frac{1}{4Dbt} \right] = \frac{\sqrt{\pi} \cdot \Gamma \left(\frac{\alpha +1}{2}\right) \cdot \left(4Dbt\right)^{\frac{1}{2}}}{2 \Gamma \left(\frac{\alpha +2}{2}\right)} - \frac{\left(4Dbt\right)^{\frac{\alpha + 2}{2}}}{\alpha +1}+ O(t^{\frac{\alpha +4}{2}}). \nonumber
\eeq
Thus for $t$ small, we have
\beq
Pr\left\lbrace t_1>t\right\rbrace \sim 1 - \frac{2 \Gamma \left(\frac{\alpha +2}{2}\right)}{(\alpha +1)\sqrt{\pi}\Gamma \left(\frac{\alpha +1}{2}\right)}(4Dbt)^{\frac{\alpha + 1}{2}}\nonumber
\eeq
and the MFAT is
\beq\label{genaral_alpha_formula}
\overline{\tau}^N
&\sim & \int_{0}^{\infty} \exp\left\lbrace - \frac{2N \Gamma \left(\frac{\alpha +2}{2}\right)}{(\alpha +1)\sqrt{\pi}\Gamma \left(\frac{\alpha +1}{2}\right)}(4Dbt)^{\frac{\alpha + 1}{2}}  \right\rbrace dt \nonumber \\
& \sim &  \left(\frac{\sqrt{\pi}(\alpha +1)\Gamma \left(\frac{\alpha +1}{2}\right)}{2N\Gamma \left(\frac{\alpha +2}{2}\right)} \right)^{\frac{2}{\alpha +1}} \cdot  \frac{1}{4Db} \cdot\Gamma \left(\frac{\alpha +3}{\alpha +1}\right).
\eeq
When $\alpha  = 0$, the initial distribution becomes $p_0 = \frac{2\sqrt{b}}{\sqrt{\pi}}\exp\left\lbrace -by^2\right\rbrace$,
we recover the same behavior as the case where the Brownian particles are uniformly distributed in $[0, y_0]$. For $\alpha = 1$, we obtain that the MFAT decays with $1/N$. For $\alpha = 2$, we obtain that MFAT decays like $\frac{1}{N^{2/3}}$.
\subsection{MFAT inside the interval $[0,a]$}
In this part, we present asymptotic computation for the MFAT to one of the extremities of an interval. We start with the solution of the diffusion equation
\beq \label{dim1interval}
\frac{\partial p(x,t)}{\partial t} &=& D \frac{\partial^2 p(x,t)}{\partial x^2} \, \hbox{ for $x>0$, $t>0$} \nonumber \\
p(x,0) &=&\delta\left(x-y\right) \,\hbox{ for $x >0$} \nonumber \\
p(0,t) &=& p(a,t) = 0\hspace{0.3 cm}\hbox{ for $t>0$}
\eeq
which is given by the infinite sum
\beq \label{pi}
p(x,t) = \frac{1}{\sqrt{4 D \pi t}} \sum_{n = -\infty}^{+ \infty}\left[\exp\left\lbrace -\frac{(x-y+2na)^2}{4Dt}\right\rbrace - \exp\left\lbrace -\frac{(x+y+2na)^2}{4Dt}\right\rbrace\right].
\eeq
{To compute the MFAT, we shall use only the first terms associated with $n = \pm 1$ \cite{Basnayake2018}.}
For an initial uniformly distributed particles in the interval $[0,a]$ of the form $p_0 (x) = \frac{1}{b} \mathbb{I}_{\left\lbrace x \in [0,b] \right\rbrace}$ with $0<b<a$, the approximated solution of equation (\ref{dim1interval}) is given by
\beq
p(x,t) &=& \int_{0}^{b} \frac{1}{b \sqrt{4Dt \pi}} \left[\exp\left\lbrace - \frac{(x-y)^2}{4Dt}\right\rbrace - \exp\left\lbrace - \frac{(x+y)^2}{4Dt}\right\rbrace + \exp\left\lbrace - \frac{(x-y-2a)^2}{4Dt}\right\rbrace \right. \nonumber \\
&-& \left. \exp\left\lbrace - \frac{(x+y-2a)^2}{4Dt}\right\rbrace + \exp\left\lbrace - \frac{(x-y+2a)^2}{4Dt}\right\rbrace - \exp\left\lbrace - \frac{(x+y+2a)^2}{4Dt}\right\rbrace \right] dy. \nonumber
\eeq
The survival probability  is
\beq
Pr\left\lbrace t_1>t\right\rbrace &=& \int_{0}^{a} p(x,t) dx =
\frac{1}{b}\left[b \cdot \mathrm{erf} \left(\frac{b}{\sqrt{4Dt}}\right) - (a+b) \cdot \mathrm{erf} \left(\frac{a+b}{\sqrt{4Dt}}\right) + 2a \cdot  \mathrm{erf} \left(\frac{a}{\sqrt{4Dt}}\right) \right.\nonumber \\
&-&(a-b) \cdot \mathrm{erf} \left(\frac{a-b}{\sqrt{4Dt}}\right)
+ (2a+b) \cdot \mathrm{erf} \left(\frac{2a+b}{\sqrt{4Dt}}\right) -4a\cdot \mathrm{erf} \left(\frac{2a}{\sqrt{4Dt}}\right) \nonumber \\
&+&(2a-b)\cdot \mathrm{erf} \left(\frac{2a-b}{\sqrt{4Dt}}\right)
- \frac{(3a-b)}{2}\cdot \mathrm{erf} \left(\frac{3a-b}{\sqrt{4Dt}}\right) -\frac{(3a+b)}{2} \cdot \mathrm{erf} \left(\frac{3a+b}{\sqrt{4Dt}}\right)\nonumber \\
&+&3a \cdot \mathrm{erf} \left(\frac{3a}{\sqrt{4Dt}}\right)
+ \frac{\sqrt{4Dt}}{\pi}\left[e^{-\frac{b^2}{4Dt}} - 1 -e^{-\frac{(a+b)^2}{4Dt}} +2e^{-\frac{a^2}{4Dt}} - e^{-\frac{(a-b)^2}{4Dt}}
+ e^{-\frac{(2a+b)^2}{4Dt}}  \right. \nonumber \\
&-& 2e^{-\frac{(2a)^2}{4Dt}} +\left. \left. e^{-\frac{(2a-b)^2}{4Dt}} - \frac{1}{2}e^{-\frac{(3a-b)^2}{4Dt}} +  e^{-\frac{(3a)^2}{4Dt}} - \frac{1}{2}e^{-\frac{(3a+b)^2}{4Dt}}  \right] \right]. \nonumber
\eeq
In the small $t$ limit, we have the approximation
\beq
S(t)=Pr\left\lbrace t_1>t\right\rbrace \sim  1 - \frac{\sqrt{4Dt}}{b\sqrt{\pi}}. \nonumber
\eeq
{Note that using equation (\ref{time dist}), we can compute the distribution of arrival times. In that case, we have
\beq \label{at1}
Pr\left\lbrace \tau^1 = t\right\rbrace &=& -\frac{d}{dt} Pr\left\lbrace \tau^1 > t\right\rbrace = N (Pr \left\lbrace t_1>t \right\rbrace)^{N-1} Pr\left\lbrace t_1 = t \right\rbrace \nonumber \\
&\sim& \frac{N \sqrt{D}}{b\sqrt{\pi \cdot t}} \exp{ \left\lbrace -\frac{\sqrt{4Dt}N}{b\sqrt{\pi}} \right\rbrace }.
\eeq}
This formula leads to the asymptotic expression for the MFAT
\beq \label{mfpt_int_1}
\overline{\tau}^N  &\sim& \frac{b^2 \pi}{2DN^2}.
\eeq
When the initial distribution intersect the right hand-side of the interval $p_0 (x) = \frac{1}{a-b} \mathbb{I}_{\left\lbrace x \in [b,a] \right\rbrace}$ with $0<b<a$, we obtain a similar expression:
\beq
\overline{\tau}^N  &\sim& \frac{(a-b)^2 \pi}{2DN^2}. \nonumber
\eeq
When the Brownian particles are initially uniformly distributed in an interval $[b,c]$ contained inside $[0,a]$, $p_0 (x) = \frac{1}{c-b} \mathbb{I}_{\left\lbrace x \in [b,c] \right\rbrace}$ with $0<b<c<a$,
the solution of the diffusion equation (\ref{dim1interval}) becomes
\beq
p(x,t) &=& \int_{b}^{c} \frac{1}{(c-b) \sqrt{4Dt \pi}}  \left[\exp\left\lbrace - \frac{(x-y)^2}{4Dt}\right\rbrace - \exp\left\lbrace - \frac{(x+y)^2}{4Dt}\right\rbrace + \exp\left\lbrace - \frac{(x-y-2a)^2}{4Dt}\right\rbrace \right. \nonumber \\
	&-& \left. \exp\left\lbrace - \frac{(x+y-2a)^2}{4Dt}\right\rbrace + \exp\left\lbrace - \frac{(x-y+2a)^2}{4Dt}\right\rbrace - \exp\left\lbrace - \frac{(x+y+2a)^2}{4Dt}\right\rbrace \right] dy \nonumber
\eeq
and the survival probability is
\beq
	Pr\left\lbrace t_1>t\right\rbrace 
	& = & \frac{1}{c-b}\left[c-b+ \frac{(\sqrt{4Dt})^3}{\sqrt{\pi}}\left[\frac{e^{-\frac{c^2}{4Dt}}}{2c^2} -\frac{e^{-\frac{b^2}{4Dt}}}{2b^2} -\frac{e^{-\frac{(c+a)^2}{4Dt}}}{2(c+a)^2} +\frac{e^{-\frac{(b+a)^2}{4Dt}}}{2(b+a)^2}
	-\frac{e^{-\frac{(a-c)^2}{4Dt}}}{2(a-c)^2}\right.\right.\nonumber \\
	&+& \frac{e^{-\frac{(c+2a)^2}{4Dt}}}{2(c+2a)^2} +\frac{e^{-\frac{(a-b)^2}{4Dt}}}{2(a-b)^2} -\frac{e^{-\frac{(b+2a)^2}{4Dt}}}{2(b+2a)^2} +\frac{e^{-\frac{(2a-c)^2}{4Dt}}}{2(2a-c)^2}
	-\frac{e^{-\frac{(2a-b)^2}{4Dt}}}{2(2a-b)^2} -\frac{e^{-\frac{(3a-c)^2}{4Dt}}}{2(3a-c)^2}\nonumber \\
	&+&\left. \left. \frac{e^{-\frac{(3a-b)^2}{4Dt}}}{2(3a-b)^2} - \frac{e^{-\frac{(c+3a)^2}{4Dt}}}{2(c+3a)^2} + \frac{e^{-\frac{(b+3a)^2}{4Dt}}}{2(b+3a)^2}  \right] \right]. \nonumber
\eeq
{The distribution for the arrival time is given by
\beq \label{at2}
Pr\left\lbrace \tau^1 = t\right\rbrace
&=& -\frac{d}{dt} [S(t)]^N \sim  -\frac{d}{dt}\left[\exp \left \lbrace- \frac{N(\sqrt{4Dt})^3 e^{-\frac{ (\min(b, a-c))^2}{4Dt}}}{2(c-b)(\min(b, a-c))^2\sqrt{\pi}} \right \rbrace \right] \nonumber \\
&\sim& \frac{N (\sqrt{4Dt})^3}{2(c-b) \cdot \min(b,a-c)\sqrt{\pi \cdot t}}
\exp{ \left\lbrace - \frac{\min^2(b,a-c)}{4Dt} \right\rbrace }\times \nonumber \\
&& \exp{ \left\lbrace -\frac{(\sqrt{4Dt})^3 N}{2(c-b)\sqrt{\pi}} \frac{ \exp{ \left\lbrace - \frac{\min^2(b,a-c)}{4Dt} \right\rbrace }}{\min^2(b,a-c)}  \right\rbrace}\left[ \frac{\min(b,a-c)}{4Dt^2} +\frac{3}{2t}\right].
\eeq
Using a Taylor expansion when $c \rightarrow b$ in the form $c = b(1+\eps)$ when $\eps\rightarrow{0}$, the survival probability $S_{\eps}(t)$ converges to the survival probability $S_{0}(t)$ in the case that the initial condition is a Dirac delta function. However, as shown above, the convergence is not uniform in time $t$ in the interval $[0,\infty[$, preventing to use this expansion to estimate the MFAT for the case of an interval. Thus to leading order, we have
\beq
\underset{c \rightarrow b}{\lim} Pr\left\lbrace t_1>t\right\rbrace	 \sim 1 - \frac{\sqrt{4Dt}}{\sqrt{\pi}}\left[
\frac{e^{-\frac{\min(b,a-c)^2}{4Dt}}}{\min(b,a-c)}\right], \nonumber
\eeq
and thus the leading order term for the asymptotic formula for $N\gg 1$ is given by
\beq \label{mfpt_int_2}
\overline{\tau}_{\eps}^N \sim \frac{\min(b,a-c)^{2}}{4D \,\,\log\left(\frac{N}{\sqrt{\pi}}\right)+A_{\eps}},
\eeq
where $A_{\eps}=A_0+\eps A_1+..$, where $A_k$  are constants. Here $\min(b,a-c)$ is the shortest distance to the absorbing boundaries. To conclude, at leading order, the MFAT when the initial distribution of particles falls inside a small interval is the same as the one obtained for a Dirac delta function where the main parameter is the minimal distance to the boundaries  of the interval where the particles are uniformly distributed.}
\subsection{MFAT  for particles initially distributed following a long tail inside the interval [0,c]}
We consider the MFAT when the initial distribution is given by $p_1 (x) = \frac{2b}{1 - e^{-bc^2}}xe^{-x^2} \mathbb{I}_{\left\lbrace x \in [0,c] \right\rbrace}$ with $a>c>0$. Then, the solution of the diffusion equation (\ref{dim1condeltadirac}) is given by the following expression
\beq
p(x,t)&=& \frac{2b}{1 - e^{-ba^2}}\int_{0}^{a} \frac{1}{\sqrt{4Dt \pi}} \left[\exp\left\lbrace - \frac{(x-y)^2}{4Dt}\right\rbrace - \exp\left\lbrace - \frac{(x+y)^2}{4Dt}\right\rbrace
+ \exp\left\lbrace - \frac{(x-y-2a)^2}{4Dt}\right\rbrace  \right. \nonumber \\
&-& \left. \exp\left\lbrace - \frac{(x+y-2a)^2}{4Dt}\right\rbrace + \exp\left\lbrace - \frac{(x-y+2a)^2}{4Dt}\right\rbrace -  \exp\left\lbrace - \frac{(x+y+2a)^2}{4Dt}\right\rbrace \right]y\cdot \exp\left\lbrace - bx^2\right\rbrace  dy. \nonumber
\eeq
For $t$ small, the survival probability can be approximated by the formula
\beq
	S(t)=Pr\left\lbrace t_1>t\right\rbrace \sim 1 - \frac{2Dbt}{1-e^{-bc^2}}.\nonumber
\eeq
{The distribution for the first arrival time in this case is
\beq \label{at3}
Pr\left\lbrace \tau^1 = t\right\rbrace
&=& -\frac{d}{dt} S^N(t) \sim  -\frac{d}{dt}\left[\exp \left \lbrace\frac{-2NDbt}{1-e^{-bc^2}} \right \rbrace \right] \nonumber \\
&=&   \frac{2bDN}{1-\exp \lbrace -bc^2 \rbrace } \exp{ \left\lbrace -\frac{2DbtN}{1-\exp \lbrace -bc^2 \rbrace } \right\rbrace }.
\eeq}
Thus, we have the asymptotic formula
\beq \label{mfpt_int_3}
\overline{\tau}^N  \sim   \frac{1-e^{-bc^2}}{2DbN}.
\eeq
\begin{figure}[http!]
\begin{center}
\includegraphics[scale = 0.55]{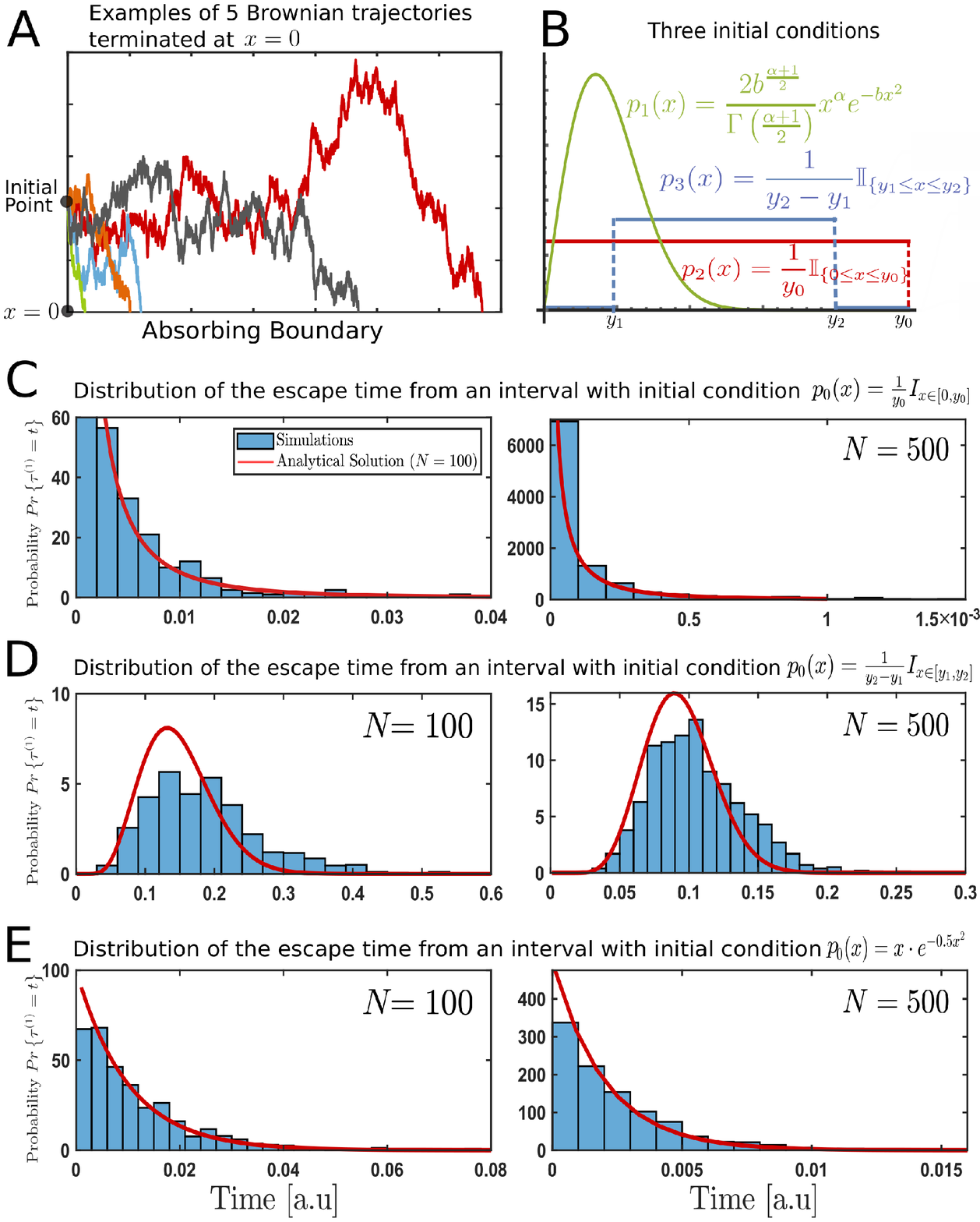}
\caption{\textbf{Arrival times for the fastest Brownian particles for various initial  distributions.} \textbf{A.} Examples of $5$ independent Brownian trajectories starting at $x=0.5$ and absorbed at $x=0$ and the fastest is green. \textbf{B.} Three initial distributions: the exponential distribution $p_1(x) = \frac{2b^{\frac{\alpha+1}{2}}}{\Gamma \left(\frac{\alpha +1}{2}\right)}x^{\alpha} e^{ -bx^2}$ and two uniform distributions $p_2(x)=\frac{1}{y_0} \mathbb{I}_{\lbrace 0 \leq x \leq y_0 \rbrace}$ and $p_3(x)=\frac{1}{y_2 - y_1} \mathbb{I}_{\lbrace y_1 \leq x \leq y_2 \rbrace}$. \textbf{C.} Distribution of the arrival time $\bar{\tau}^N$:  analytical (equation (\ref{at1})) in red vs stochastic simulations (blue histogram) for particles distributed with respect to $p_0 (x) = \frac{1}{b} \mathbb{I}_{\left\lbrace x \in [0,b] \right\rbrace}$ for $0<b<a$ with $b=4$ and $a = 5$ for $N=100$ (left) and $N=500$ (right) with $1000$ runs. \textbf{D.} Distribution of the arrival time $\bar{\tau}^N$:  analytical (equation (\ref{at2})) in red vs stochastic simulations (blue) for particles distributed with respect to $p_0 (x) = \frac{1}{c-b} \mathbb{I}_{\left\lbrace x \in [b,c] \right\rbrace}$ with $0<b<c<a$, $b=1$ and $c = 4$.\textbf{E.} Distribution of the arrival time $\bar{\tau}^N$:  analytical (equation (\ref{at3})) in red vs stochastic simulations (blue) for particles distributed with respect to $p_1 (x) = \frac{2b}{1 - e^{-bc^2}}xe^{-x^2} \mathbb{I}_{\left\lbrace x \in [0,c] \right\rbrace}$ with $a>c>0$ with $b = 0.5$, $\alpha = 1$ and $c=4$.}
\label{graph1}
\end{center}
\end{figure}
{We decided to compare the asymptotic distribution we obtained with stochastic simulations (see Appendix for the description of the algorithm). We generated  trajectories before the reach the origin and selected the fastest (green in Fig. \ref{graph1}A).  We chose several  initial distributions (Fig. \ref{graph1}B) and we compare the histogram of arrival time for the fastest with the analytical expression in Fig. \ref{graph1}C, D and E when the domain of simulation is the interval $[0,a]$. We found a very good agreement between the analytical distributions and the empirical histogram  for the arrival times of the fastest for the the three initial distributions associated to the pdf of the fastest given by expressions (\ref{at1}), (\ref{at2}), and (\ref{at3}) respectively.}
\begin{figure}[http!]
\begin{center}
\includegraphics[scale = 0.48]{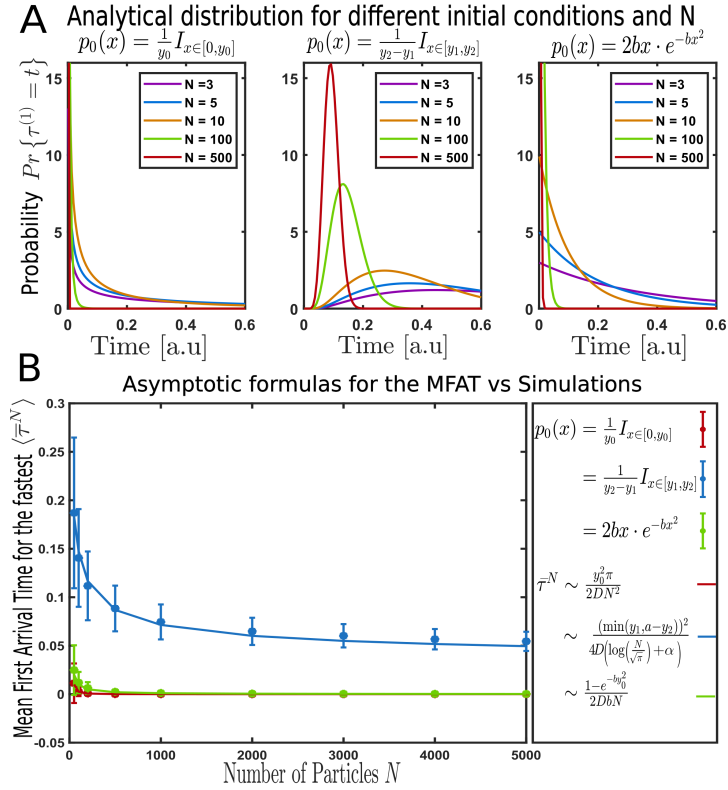}
\caption{\textbf{Mean fastest arrival time vs the number of particles $N$.} \textbf{A.} Probability $Pr\left\lbrace \tau^1 = t\right\rbrace$ of arrival times in an interval computed from equations (\ref{at1}), (\ref{at2}), and (\ref{at3}) for a total number $N = [3, 5, 10, 100,500]$ and the three initial distributions, presented in Fig. \ref{graph1}B). \textbf{B.} MFAT  vs $N$ comparing stochastic simulations (colored disks) and the asymptotic formulas (continuous lines). The asymptotic expression for the MFAT  when the particles are initially uniformly distributed in $[y_1, y_2]$ (blue curve) is of the form $\frac{y_1^2}{4D\left(\mathrm{log}\left(\frac{N}{\sqrt{\pi}}\right)+\alpha \right)}$. An optimal fit gives $\alpha = -0.3075$. Parameters of the simulations are described in Fig. \ref{graph1} with 1000 runs.}
\label{graph2}
\end{center}
\end{figure}
\\
{We plotted the analytical pdfs for the shortest arrival time (\ref{at1}), (\ref{at2}), (\ref{at3}) and for various values of $N$ (Fig. \ref{graph2}A).  The MFAT decreases with the number of particles (Fig. \ref{graph2}B) as predicted by equations (\ref{mfpt_int_1}) red, (\ref{mfpt_int_2}) blue and (\ref{mfpt_int_3}) green.  Note that in this case, we used $b=1$ and $c = 3$, for which the distribution (\ref{at2}) and the MFAT (\ref{mfpt_int_2}) show a good agreement with the stochastic simulations (Fig. \ref{graph1}D and Fig. \ref{graph2}B in blue, respectively).}
We summarized in the next table the main asymptotic formulas associated with different initial conditions.
\vspace{0.5 cm}
\begin{tabular}{ |p{3.3cm}|p{3.1cm}||p{3.9cm}|p{3cm}|  }
	\hline
	\multicolumn{4}{|c|}{Asymptotics formula for the MFAT} \\
	\hline
	Initial Distribution & $\Omega = \mathbf{R}_{+}$ & Initial Distribution & $\Omega = [0,a]$ \\
	\hline
	$\scriptstyle p_0(x) = \delta(x-x_0)$ & $\scriptstyle \overline{\tau}^N  \sim \frac{x_0^2}{4D \cdot \mathrm{log} \left(\frac{N}{\sqrt{\pi}} \right)}$ & $\scriptstyle p_0(x) = \delta(x-x_0)$ & $\scriptstyle  \overline{\tau}^N  \sim \frac{x_0^2}{16D \cdot \mathrm{log} \left(\frac{2N}{\sqrt{\pi}} \right)}$ \\
	\hline
	$\scriptstyle p_0(x) = \frac{1}{y_0} \mathbb{I}_{\left\lbrace x \in [0,y_0] \right\rbrace}$ & $\scriptstyle \overline{\tau}^N  \sim  \frac{y_0^2 \pi}{2DN^2}$ &$\scriptstyle p_0(x) = \frac{1}{y_0} \mathbb{I}_{\left\lbrace x \in [0,y_0] \right\rbrace}$ & $\scriptstyle \overline{\tau}^N \sim \frac{y_0^2 \pi}{2DN^2}$ \\
	\hline
	$\scriptstyle p_0(x) = \frac{1}{y_2-y_1} \mathbb{I}_{\left\lbrace x \in [y_1,y_2] \right\rbrace}$  & $ \scriptstyle \overline{\tau}^N  \sim \frac{y_1 ^{2}}{4D \cdot \mathrm{log} \left(\frac{N}{\sqrt{\pi}}\right)}$ & $\scriptstyle p_0(x) = \frac{1}{y_2-y_1} \mathbb{I}_{\left\lbrace x \in [y_1,y_2] \right\rbrace}$ & $\scriptstyle \overline{\tau}^N  \sim \frac{(\min(y_1, a-y_2))^2}{4D \cdot \mathrm{log} \left(\frac{N}{\sqrt{\pi}}\right)}$ \\
	\hline
	$\scriptstyle p_0(x) = 2bx e^{ -bx^2}$ & $\scriptstyle \overline{\tau}^N  \sim \frac{1}{2DbN}$ & $\scriptstyle p_0(x) = \frac{2b}{1 - e^{-bc^2}}xe^{-x^2} \mathbb{I}_{\left\lbrace x \in [0,c] \right\rbrace}$ & $\scriptstyle \overline{\tau}^N  \sim\frac{1-e^{-bc^2}}{2DbN}$\\
	\hline
	$\scriptstyle p_0(x) = \frac{4b^{\frac{3}{2}}}{\sqrt{\pi}}x^2 e^{ -bx^2}$ & $\scriptstyle \overline{\tau}^N  \sim \frac{\pi ^{\frac{2}{3}}}{4Db(2N)^{\frac{2}{3}}}\cdot \Gamma\left(\frac{5}{3}\right)$ & -& -\\
	\hline
\end{tabular}
To conclude, when there are $N$ i.i.d. Brownian particles initially uniformly distributed in an interval that does not contain the escape points, either the real semi-axis or a bounded interval, then the MFAT has a similar decay with $N$ for a Dirac delta function or a locally constant initial distribution.
{
\section{MFAT in dim 2}
\subsection{MFAT for a uniform initial distribution }
We study here $N$ i.i.d. Brownian particles in a bounded domain in two dimension $\Omega \subset \mathbb{R}^ 2$  (Fig. \ref{graph4}). The particles are initially uniformly distributed in the region
\beq
\Omega ^* = \left\lbrace B_{r_2}(A) \setminus B_{r_1}(A) : \theta_1 \leq \theta \leq \theta_1 + \theta_2 \right \rbrace,
\eeq
where $ B_{r}(A)$ is a disk of radius r centered at A. They can bind to a single small absorbing arc $\partial \Omega_a$ in the boundary $\partial \Omega$ of $\Omega$ of length $2 \eps$ and centered at a point $A$.
\begin{figure}[http!]
\begin{center}
\includegraphics[scale = 0.75]{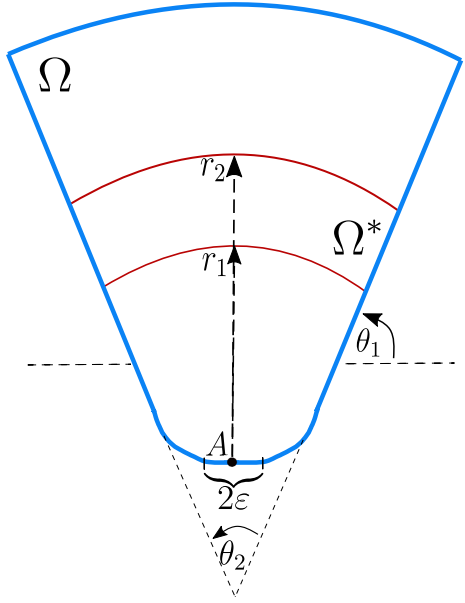}
\caption{\textbf{Schematic of the two-dimensional domain to study  the MFAT to a small arc.} The bounded domain $\Omega$ is delimited by the blue curve. The initial distribution of the Brownian particles is given by $p(x,0) = \frac{1}{A(\Omega^*)}\mathbb{I}_{\left \lbrace x \in \Omega^* \right \rbrace}$, where the region $\Omega ^* = \left\lbrace B_{r_2}(A) \setminus B_{r_1}(A) : \theta_1 \leq \theta \leq \theta_1 + \theta_2 \right \rbrace$ is delimited by the red curves.}
\label{graph4}
\end{center}
\end{figure}
Here we consider the initial distribution of i.i.d Brownian particles
\beq
p(x,0) = \frac{1}{A(\Omega^*)}\mathbb{I}_{\left \lbrace x \in \Omega^* \right \rbrace}
\eeq
where the area of the region is $A(\Omega^*) = \frac{(r_2^2-r_1^2)\theta_2}{2}$. The solution of the diffusion equation with a general initial condition is the convolution of the elementary solution for the Dirac delta function with the initial condition $\tilde{p}(x,t)$:
\beq
p(x,t) = \int_{x} \tilde{p}(y-x,t) \ast_x p(y,0)dy.
\eeq
Using the asymptotic solution computed in \cite{Basnayake2018} for dimension two, we have
\beq
S(t) &=& Pr\left\lbrace t_1>t\right\rbrace =\frac{1}{A(\Omega^*)}\underset{\Omega^*}{\int} \left(1-\frac{\sqrt{2}D\pi t}{2 \log \left(\frac{1}{\eps} \right)} \frac{\exp \left \lbrace - \frac{|A-y|^ 2}{4Dt}\right \rbrace}{|A-y|^ 2}\right) dy \nonumber \\
&=& 1- \frac{\sqrt{2}D\pi t}{2 \log \left(\frac{1}{\eps} \right)A(\Omega^*)}  \underset{\Omega^*-A}{\int} \frac{\exp \left \lbrace - \frac{|z|^ 2}{4Dt}\right \rbrace}{|z|^ 2} dy. \nonumber
\eeq
Note now, $\Omega^*-A = \left\lbrace B_{r_2}(0) \setminus B_{r_1}(0) : \theta_1 \leq \theta \leq \theta_1 + \theta_2 \right \rbrace$, then $|z|^2 = r^2$, where $r$ is the distance to the point A chosen as the origin of coordinates. Then we can rewrite the integral above as
\beq
S(t)&=& 1- \frac{\sqrt{2}D\pi t}{2 \log \left(\frac{1}{\eps} \right)A(\Omega^*)}  \int_{r_1}^{r_2} \int_{\theta_1}^{\theta_1 + \theta_2} \frac{\exp \left \lbrace - \frac{r^ 2}{4Dt}\right \rbrace}{r^ 2} r d \theta dr \nonumber \\
&=& 1- \frac{\sqrt{2}D\pi t}{2 \log \left(\frac{1}{\eps} \right)\left(r_2^2 - r_1^2\right)}  \left( \mathrm{ei}\left(-\frac{r_2^ 2}{4Dt}\right)-\mathrm{ei}\left(-\frac{r_1^ 2}{4Dt}\right)\right),
\eeq
where $\mathrm{ei(x)} = - \int_{-x}^{\infty} \frac{e^{-t}}{t}dt$ is the exponential integral. For $t$ small,  we expand this function and we get
\beq
S(t)\sim 1- \frac{\sqrt{2}D\pi t}{2 \log \left(\frac{1}{\eps} \right)\left(r_2^2 - r_1^2\right)}  \left( \frac{4Dt}{r_1^2} e^{-\frac{r_1^2}{4Dt}}- \frac{4Dt}{r_2^2} e^{-\frac{r_2^2}{4Dt}}\right).
\eeq
Scaling $r_2 = (1+\gamma)r_1$ and  making an expansion when $\gamma \rightarrow 0$, we get
\beq
S(t) \sim 1- \frac{\sqrt{2}\pi (4Dt)e^{-\frac{r_1^2}{4Dt}}}{8 \log \left(\frac{1}{\eps} \right)r_1^2} \left(1 + \frac{4Dt}{r_1^2} -2\gamma \left(1 + \frac{r_1^2}{4Dt}\right) + O(\gamma^ 2)+ O\left(\frac{\gamma}{t} \right) \right).
\eeq
To conclude, when $\gamma \rightarrow 0$, the survival probability $S(t)$ is recovered for the case with the Dirac delta function at the points with $r=r_1$ as the initial condition. Thus to leading order, using that
\beq
\underset{r_2 \rightarrow r_1}{\lim} Pr\left\lbrace t_1>t\right\rbrace	 \sim 1 -\frac{\sqrt{2}\pi (4Dt)e^{-\frac{r_1^2}{4Dt}}}{8 \log \left(\frac{1}{\eps} \right)r_1^2} , \nonumber
\eeq
we obtain to leading order the asymptotic formula for $N\gg 1$
\beq
\overline{\tau}_{\eps}^N \approx \frac{r_1^{2}}{4D \,\,\log\left(\frac{\sqrt{2} \pi N}{8\log \left(\frac{1}{\eps} \right)}\right)+A_{\eps}},
\eeq
where $A_{\eps}=A_0+\eps A_1+..$, where $A_k$  are constants as before and $r_1$ is the shortest distance to the absorbing boundary. The MFAT for this case has a similar behavior as for the Dirac delta function.
\subsection{MFAT in two dimensions for an initial distribution asymptotically intersecting the target site}
We study here the  consequence of an initial distribution asymptotically intersecting the target site. For that goal, we consider the algebraic distribution modulated by a global exponential
\beq
p_1(x) = K |x-A|^{\alpha}e^{-b|x-A|^2}.
\eeq
The function $p_1(x)$ is the initial distribution for the diffusion equation (\ref{dim1condeltadirac}), where the normalization constant is approximated on most of the domain where we added the small triangle at the summit between the dashed and the blue lines in Fig. \ref{graph4} (the area of which is $O(\eps^2$). Thus we approximate  normalization constant is
\beq
K_{\alpha}^{-1} &\approx & \int_{\Omega} |x-A|^{\alpha}e^{-b|x-A|^2} dx =\int_{0}^{R} \int_{\theta_1}^{\theta_1 + \theta_2} r^{\alpha}e^{-br^2} r d \theta dr, \nonumber
\eeq
where $R$ is the largest radius of the circular sector centered in $A$ that can be inscribed in $\Omega$.
This leads to the expression
\beq
K_{\alpha}^{-1} &=& \frac{\theta_2}{2b^{\frac{2+\alpha}{2}}} \left(\Gamma \left(1+\frac{\alpha}{2}\right) - \Gamma \left(1+\frac{\alpha}{2}, bR^2\right) \right),
\eeq
where $\Gamma(z)$ is the Gamma function and $\Gamma(z,t)$ is the incomplete Gamma function. We can now estimate the survival probability
\beq
S(t) &=&Pr\left\lbrace t_1>t\right\rbrace= \int_{\Omega} \int_{\Omega} \tilde{p}(x,t) K_{\alpha}^{-1}|y-A|^{\alpha}e^{-b|y-A|^2}dy dx \nonumber \\
&=& \int_{\Omega} K_{\alpha}^{-1} \left(1-\frac{\sqrt{2}D\pi t}{2 \log \left(\frac{1}{\eps} \right)} \frac{\exp \left \lbrace - \frac{|A-y|^ 2}{4Dt}\right \rbrace}{|A-y|^ 2}\right)|y-A|^{\alpha}e^{-b|y-A|^2}dy \nonumber \\
&=& 1 -\frac{\sqrt{2}D\pi t}{2 \log \left(\frac{1}{\eps}\right) K_{\alpha}}\int_{\Omega} |y-A|^{\alpha-2}e^{-\frac{4Dbt+1}{4Dt}|y-A|^2}dy \nonumber \\
&=& 1 -\frac{\sqrt{2}D\pi t \theta_2}{2 \log \left(\frac{1}{\eps}\right) K_{\alpha}}\int_{0}^R r^{\alpha-1}e^{-\frac{4Dbt+1}{4Dt}r^2}dr \nonumber \\
&=& 1 -\frac{\sqrt{2}\pi (4Dbt)^{\frac{\alpha + 2}{2}}\left(\Gamma \left(\frac{\alpha}{2}\right) - \Gamma \left(\frac{\alpha}{2}, \frac{4Dbt+1}{4Dt} R^2\right) \right)}{8 \log \left(\frac{1}{\eps}\right)(4Dbt+1)^{\frac{\alpha}{2}} \left(\Gamma \left(1+\frac{\alpha}{2}\right) - \Gamma \left(1+\frac{\alpha}{2}, bR^2\right) \right)}. \nonumber
\eeq
Thus for $t$ small, we have
\beq
S(t) \sim  1 -\frac{\sqrt{2}\pi \Gamma \left(\frac{\alpha}{2}\right)  }{8 \log \left(\frac{1}{\eps}\right) \left(\Gamma \left(1+\frac{\alpha}{2}\right) - \Gamma \left(1+\frac{\alpha}{2}, bR^2\right) \right)}(4Dbt)^{\frac{\alpha + 2}{2}}, \nonumber
\eeq
and the MFAT is
\beq
\overline{\tau}^N
&\sim & \int_{0}^{\infty} \exp\left\lbrace - \frac{\sqrt{2}N\pi \Gamma \left(\frac{\alpha}{2}\right)  }{8 \log \left(\frac{1}{\eps}\right) \left(\Gamma \left(\frac{\alpha +2}{2}\right) - \Gamma \left(\frac{\alpha+2}{2}, bR^2\right) \right)}(4Dbt)^{\frac{\alpha + 2}{2}}  \right\rbrace dt \nonumber \\
& \sim &  \left(\frac{8 \log \left(\frac{1}{\eps}\right) \left(\Gamma \left(\frac{\alpha +2}{2}\right) - \Gamma \left(\frac{\alpha+2}{2}, bR^2\right) \right)}{\sqrt{2}N\pi \Gamma \left(\frac{\alpha}{2}\right)} \right)^{\frac{2}{\alpha +2}} \cdot  \frac{1}{4Db} \cdot\Gamma \left(\frac{\alpha +4}{\alpha +2}\right).
\eeq
We can rewrite this formula as
\beq
\bar{\tau}^{N}\sim  \frac{C_{\alpha, \Omega}}{N^{\frac{2}{\alpha +2}}}
 \frac{1}{4Db} \text{   for $N\gg 1$},
\eeq
where
\beq
C_{\alpha, \Omega}= \Gamma \left(\frac{\alpha +4}{\alpha +2}\right) \left(\frac{8 \log \left(\frac{1}{\eps}\right) \left(\Gamma \left(\frac{\alpha +2}{2}\right) - \Gamma \left(\frac{\alpha+2}{2}, bR^2\right) \right)}{\sqrt{2}\pi \Gamma \left(\frac{\alpha}{2}\right)} \right)^{\frac{2}{\alpha +2}}.
\eeq
To conclude, we propose that the present formula could be generalized to any dimension $d$, that would lead to a spectrum of possible decay  of the MFAT with respect to the variable $N$ depending in the algebraic decay $x^alpha$ of the initial distribution at the target located at 0:
\beq
\bar{\tau}^{N}\sim  \frac{C_{\alpha, \Omega}}{N^{\frac{2}{\alpha +d}}}
 \frac{1}{4Db} \text{   for $N\gg 1$},
\eeq
where $d$ is the dimension of $\Omega$.}
\section{Effect of a constant drift on extreme arrival} \label{sec_exp_2}
In  half-a-line, we consider now the first arrival time of the $N$ independent processes $(X_1(t),..X_N(t))$, solution of
\beq
\dot{X}_k =-a+\sqrt{2D} \dot{w}_k, \nonumber
\eeq
where $a$ is a constant velocity and $D$ the diffusion coefficient.
\subsection{Effect of a constant drift when the initial distribution is a Dirac delta function} \label{s:secdd}
\begin{figure}[http!]
\begin{center}
\includegraphics[scale = 0.5]{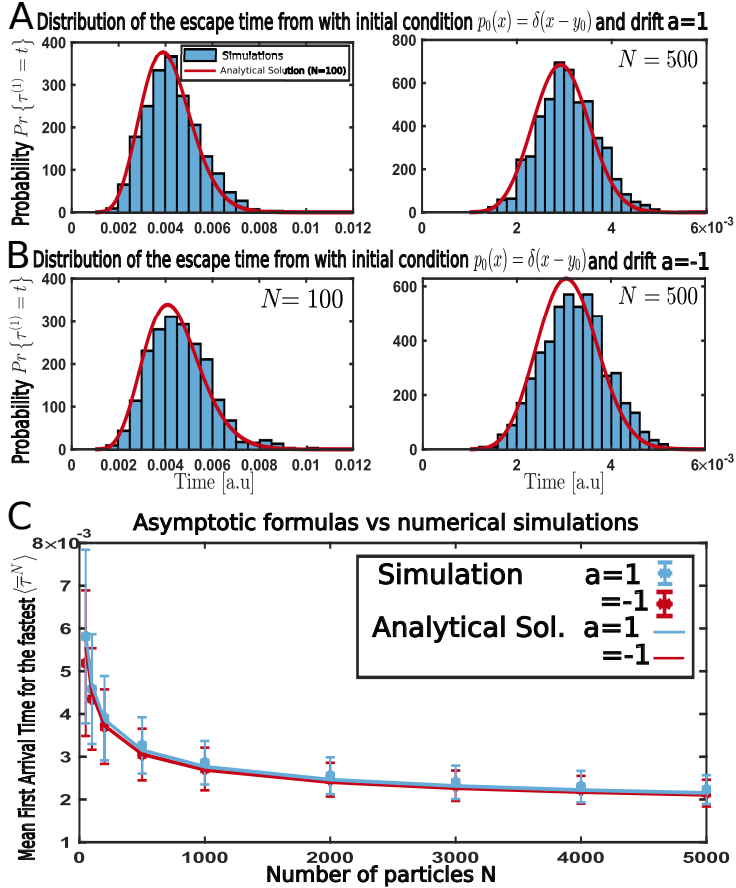}
\caption{\textbf{Mean fastest arrival time vs the number of particles $N$ with a drift.} \textbf{A.} Distribution of the arrival time $\bar{\tau}^N$:  analytical represented by equation (\ref{at4}) (in red) vs stochastic simulations (blue histogram) for particles distributed with respect to $p_0 (x) = \delta(x-y_0)$  with $y_0 = 0.25$ and a drift $a=1$ for $N=500$ (left) and $N=500$ (right) with $1000$ runs. \textbf{B.} Distribution of the arrival time $\bar{\tau}^N$:  analytical expression (\ref{at4}) (in red) vs stochastic simulations (blue histogram) with drift $a=-1$. \textbf{C.} MFAT  vs $N$ obtained from stochastic simulations (colored disks) and the asymptotic formulas (continuous lines) with $y_0 = 0.25$ and 1000 runs. }
\label{graph3}
\end{center}
\end{figure}

The Fokker-Planck Equation (FPE) is
\beq \label{eqdrift}
\frac{\partial p(x,t)}{\partial t} &=& D \frac{\partial^2 p(x,t)}{\partial x^2} +a \frac{\partial p(x,t)}{\partial x}\, \hbox{ for } x>0 , t>0  \\
p(x,0) &=& p_0(x)=\delta(x-y) \,  \nonumber \\
p(0,t) &=&0. \nonumber
\eeq
To solve equation (\ref{eqdrift}), we change the variable $p=q\exp \left\lbrace \frac{-a x+a^2t/2}{2D} \right\rbrace$, so that $q$ satisfies the diffusion equation $\frac{\partial q(x,t)}{\partial t} = D \frac{\partial^2 q(x,t)}{\partial x^2}$. Thus the solution is given by equation (\ref{p}) and
\beq
p(x,t)=\frac{\exp\left\lbrace \frac{ -a (x-y)-a^2t/2}{2D}\right\rbrace}{\sqrt{4 D \pi t}}\left[\exp\left\lbrace -\frac{(x-y)^2}{4Dt}\right\rbrace - \exp\left\lbrace -\frac{(x+y)^2}{4Dt}\right\rbrace\right]. \nonumber
\eeq
The extreme escape time is given by relation (\ref{mfpt}) \beq \label{mfpt2}
\overline{\tau}^{N} = \int_{0}^{\infty} S(t)^{N} dt,
\eeq
where the survival probability is
\beq
S(t)&=& \int_{0}^{\infty} \frac{\exp \left\lbrace \frac{-a (x-y)+a^2t/2}{2D}\right\rbrace}{\sqrt{4 D \pi t}}\left[\exp\left\lbrace -\frac{(x-y)^2}{4Dt}\right\rbrace - \exp\left\lbrace -\frac{(x+y)^2}{4Dt}\right\rbrace\right] dx\nonumber\\
&=& S_1(t)-S_2(t). \nonumber
\eeq
with
\beqq
S_1(t)=\frac{1}{\sqrt{\pi}}\int_{ \frac{-y+at}{\sqrt{4 D \pi t}}}^{\infty} \exp (-u^2)du \hbox{ and }
S_2(t)=\frac{1}{\sqrt{\pi}}\exp \left\lbrace \frac{ay}{D} \right\rbrace \int_{ \frac{y+at}{\sqrt{4 D \pi t}}}^{\infty} \exp (-u^2)du.
\eeqq
We re-write the sum $ S(t)=H_1(t)+H_2(t),$
\beq
H_1(t)&=&\frac{1}{2} \left( \mathrm{erfc} \left( \frac{-y+at}{\sqrt{4 D \pi t}}\right)-\mathrm{erfc}\left(\frac{y+at}{\sqrt{4 D \pi t}}\right)\right)\nonumber\\
H_2(t)&=&\frac{1}{2}\left(1-\exp \left\lbrace\frac{ay}{D}\right\rbrace\right)  \mathrm{erfc}\left(\frac{y+at}{\sqrt{4 D \pi t}}\right).\nonumber
\eeq
We obtain for short-time asymptotic,
\beq
H_1(t)&\sim& 1-\sqrt{4 D t} \frac{e^{-\frac{y^2}{(4Dt)}}}{y \sqrt{\pi}} \nonumber\\
H_2(t)&\sim& \left(1-\exp \left\lbrace \frac{ay}{D}\right\rbrace\right) \sqrt{ D t} \frac{e^{-\frac{y^2}{(4Dt)}}}{y \sqrt{\pi}}.\nonumber
\eeq
Thus, $S(t)=1- \sqrt{D t} \left( 1+ \exp \left\lbrace \frac{ay}{D}\right\rbrace \right) \frac{e^{-\frac{y^2}{(4Dt)}}}{y \sqrt{\pi}}$ leading to,
{
\beq \label{at4}
Pr\left\lbrace \tau^1 = t\right\rbrace
&=& -\frac{d}{dt} S(t) \sim  -\frac{d}{dt}\left[\exp \left\lbrace- N \sqrt{4 D t}  \frac{\left( 1+ \exp \left\lbrace \frac{ay}{D}\right\rbrace \right)}{2}  \frac{e^{-\frac{y^2}{(4Dt)}}}{y \sqrt{\pi}}\right \rbrace \right] \nonumber \\
&=&  \frac{ N\sqrt{4 D t} \frac{\left( 1+ \exp \left\lbrace \frac{ay}{D}\right\rbrace \right)}{2} }{y \sqrt{\pi}}\exp \left\lbrace- N \sqrt{4 D t}  \frac{\left( 1+ \exp \left\lbrace \frac{ay}{D}\right\rbrace \right)}{2}  \frac{e^{-\frac{y^2}{(4Dt)}}}{y \sqrt{\pi}}\right \rbrace \nonumber \\ &\times& e^{-\frac{y^2}{(4Dt)}} \left[ \frac{1}{2t} + \frac{y^2}{(4Dt^2)} \right]
\eeq
and }
\beq \label{mfpt3}
\overline{\tau}^{N} = \int_{0}^{\infty} S(t)^{N} dt \sim \int_{0}^{\infty} \exp \left\lbrace- N \sqrt{4 D t}  \frac{\left( 1+ \exp \left\lbrace \frac{ay}{D}\right\rbrace \right)}{2}  \frac{e^{-\frac{y^2}{(4Dt)}}}{y \sqrt{\pi}}\right \rbrace dt.
\eeq
Using the asymptotic computation of \cite{Basnayake2018}, we obtain
\beq \label{mfptnew}
\overline{\tau}^{N} \sim
\frac{y^2}{4D \log\left(N \frac{\left( 1 + \exp \left\lbrace\frac{ay}{D}\right\rbrace \right)}{2\sqrt{\pi}}\right)}.
\eeq
To conclude, in the large $N$ limit, adding a negative drift $a<0$, leads to a small increase in the extreme arrival time. Formula (\ref{mfptnew}) reveals how adding a drift can be equivalent to reducing the number of initial particles by a factor $\exp (\frac{ay}{D})<1$.
\\
{Finally, we tested the quality of the analytical approximation of the pdfs  with the stochastic simulations for the shortest arrival time (\ref{at4}) with drifts $a= 1$ and $a=-1$ respectively and for different values of $N$ (Fig. \ref{graph3}A and B).  The MFAT decreases with $N$ (Fig. \ref{graph3}C) according to equation (\ref{mfpt3}) for $a=-1$ in red and $a= 1$ in blue.}
\subsection{Effect of a constant drift for a uniform initial distribution}\label{s:seuu}
When the particles are initially uniformly distributed in a small portion of the ray, $[0;y_0]$, the initial condition is given by  formula $p(x,0) = \frac{1}{y_0} \mathbb{I}_{\left[x \in [0,y_0] \right]}$, and the solution of the diffusion equation with a drift (\ref{eqdrift}) is given by
\beq
p(x,t) &=& e^{\frac{\ds -ax-a^2t/2}{\ds 2D}} \int_0^
{y_0}  \frac{\exp \left\lbrace\frac{ay}{2D}\right\rbrace}{\sqrt{4 \pi D t}y_0} \left[\exp \left \lbrace-\frac{(x-y)^2}{4Dt}\right\rbrace-\exp \left \lbrace-\frac{(x+y)^2}{4Dt}\right\rbrace\right]dy \nonumber \\
&=& \frac{1}{2y_0}\left[\mathrm{erfc} \left(\frac{x-y_0 +a t}{\sqrt{4Dt}}\right)-\mathrm{erfc} \left(\frac{x +a t}{\sqrt{4Dt}}\right) \right.-\left. e^{-\frac{ax}{D}}\left(\mathrm{erfc} \left(\frac{x-a t}{\sqrt{4Dt}}\right)-\mathrm{erfc} \left(\frac{x+y_0 -a t}{\sqrt{4Dt}}\right)\right)\right]. \nonumber
\eeq
Then the survival probability is
\beq
S(t) &=& \frac{1}{2y_0} \int_{0}^{\infty}\left( \mathrm{erfc}\left(\frac{x-y_0 +a t}{\sqrt{4Dt}}\right)-\mathrm{erfc}\left(\frac{x +a t}{\sqrt{4Dt}}\right) \right) dx \nonumber \\
&-& \frac{1}{2y_0} \int_{0}^{\infty}e^{-\frac{ax}{D}} \left( \mathrm{erfc}\left(\frac{x-a t}{\sqrt{4Dt}}\right)-\mathrm{erfc}\left(\frac{x -a t + y_0}{\sqrt{4Dt}}\right) \right) dx \nonumber \\
&=& S_1(t)+S_2(t). \nonumber
\eeq
Using the change of variable $ u = \frac{x}{\sqrt{4D \pi t}}$ we can obtain,
\beq
S_1(t) &=& \frac{1}{2y_0} \left(-a t \cdot \mathrm{erf} \left(\frac{a t }{\sqrt{4Dt}}\right) + \frac{\sqrt{4Dt}}{\sqrt{\pi}} e^{- \frac{(y_0 - a t)^2}{4Dt}} - y_0 \cdot \mathrm{erf} \left(\frac{at-y_0}{\sqrt{4Dt}}\right)\right. \nonumber \\
&-& \left. \frac{\sqrt{4Dt}}{\sqrt{\pi}} e^{-\frac{a^2t^2}{4Dt}} +y_0 + a t\cdot  \mathrm{erf}\left(\frac{at - y_0}{\sqrt{4Dt}}\right) \right).\nonumber
\eeq
Then, using the change of variable $u  =\frac{x-at}{\sqrt{4Dt}}$ and integrating by parts, we obtain
\beq
S_2(t) &=& \frac{D}{2ay_0}\left(-2 \cdot \mathrm{erf}\left(\frac{at}{\sqrt{4Dt}}\right) +\mathrm{erfc}\left(\frac{y_0-at}{\sqrt{4Dt}}\right)  -  \mathrm{erfc}\left(\frac{y_0+a t }{\sqrt{4Dt}}\right)e^{\frac{ay_0}{D}}\right).\nonumber
\eeq
Thus, making the asymptotic for $t$ small, we have
\beq
S(t) \sim 1 - \frac{\sqrt{4Dt}}{\sqrt{\pi} y_0} \nonumber
\eeq
leading to
\beq
\overline{\tau}^{N} =
\frac{y_0^2 \pi}{2D N^ 2}.\nonumber
\eeq
This result is the one obtained in the case of no drift, showing that the drift does not affect the extreme arrival time when the absorbing point overlap with the interval where the particles are initially uniformly distributed.
\subsection{Effect of a constant drift for a uniform initial distribution not intersecting the target}\label{s:seuuni}
When the particles are uniformly distributed in the interval $[y_1, y_2]$ with $y_2>y_1>0$, the initial distribution is of the form $p_0(x)=\frac{1}{y_2 - y_1} \mathbb{I}_{\left[x \in [y_1,y_2] \right]}$. The solution of the diffusion equation \ref{eqdrift} is given by
\beq
p(x,t) &=& e^{\frac{\ds -ax-a^2t/2}{\ds 2D}} \int_{y_1}^
{y_2}  \frac{\exp \left\lbrace\frac{ay}{2D}\right\rbrace}{\sqrt{4 \pi D t}(y_2 - y_1)} \left[\exp \left \lbrace-\frac{(x-y)^2}{4Dt}\right\rbrace-\exp \left \lbrace-\frac{(x+y)^2}{4Dt}\right\rbrace\right]dy \nonumber \\
&=& \frac{1}{2(y_2 - y_1)}\left[\mathrm{erfc}\left(\frac{x-y_2 +a t}{\sqrt{4Dt}}\right)- \mathrm{erfc}\left(\frac{x-y_1 +a t}{\sqrt{4Dt}}\right) \right.\nonumber \\
&-& \left. e^{-\frac{ax}{D}}\left(\mathrm{erfc}\left(\frac{x+y_1-a t}{\sqrt{4Dt}}\right)-\mathrm{erfc}\left(\frac{x+y_2 -a t}{\sqrt{4Dt}}\right)\right)\right],\nonumber
\eeq
and we can compute the survival probability as
\beq
S(t) &=& \frac{1}{2(y_2-y_1)} \int_{0}^{\infty}\left( \mathrm{erfc} \left(\frac{x-y_2 +a t}{\sqrt{4Dt}}\right)-\mathrm{erfc} \left(\frac{x-y_1 +a t}{\sqrt{4Dt}}\right) \right) dx \nonumber \\
&-& \frac{1}{2(y_2-y_1)} \int_{0}^{\infty}e^{-\frac{ax}{D}} \left( \mathrm{erfc} \left(\frac{x+y_2 -a t}{\sqrt{4Dt}}\right)-\mathrm{erfc} \left(\frac{x +y_1 -a t}{\sqrt{4Dt}}\right) \right) dx \nonumber \\
&=& S_1(t)+S_2(t).\nonumber
\eeq
And, proceeding as before, we obtain
\beq
S_1(t) &=& \frac{1}{2(y_2 - y_1)}\left(2(y_2 - y_1) -(y_2 -at) \mathrm{erfc} \left(\frac{y_2 -at}{\sqrt{4Dt}}\right)
+(y_1-at)\mathrm{erfc}\left(\frac{y_1-at}{\sqrt{4Dt}}\right)  \right. \nonumber \\
&+& \frac{\sqrt{4Dt}}{\sqrt{\pi}}\exp \left\lbrace-\frac{(y_2-at)^2}{4Dt}\right\rbrace\left. -\frac{\sqrt{4Dt}}{\sqrt{\pi}}\exp \left\lbrace-\frac{(y_1-at)^2}{4Dt}\right\rbrace\right)\nonumber
\eeq
and
\beq
S_2(t) &=& \frac{-D}{2a(y_2 - y_1)}\left(\mathrm{erfc} \left(\frac{y_1-at}{\sqrt{4Dt}}\right) -\mathrm{erfc} \left(\frac{y_2-at}{\sqrt{4Dt}}\right) \right. \nonumber \\
&-& \left. e^{\frac{ay_1}{D}} \mathrm{erfc} \left(\frac{y_1+a t }{\sqrt{4Dt}}\right) + e^{\frac{ay_2}{D}} \mathrm{erfc} \left(\frac{y_2+a t }{\sqrt{4Dt}}\right)\right).\nonumber
\eeq
{The asymptotics for the short-time is given by
\beq \label{Sp_for_[y_1, y_2]_with_a_drift}
S(t) &\sim& 1 - \frac{D\sqrt{Dt}}{a\left(y_2-y_1\right)\sqrt{\pi}} \left[ \frac{\exp \left\lbrace -\frac{(y_1-at)^2}{4Dt}\right\rbrace}{y_1-at} -\frac{\exp \left\lbrace -\frac{(y_2-at)^2}{4Dt}\right\rbrace}{y_2-at} \right. \nonumber \\
&+& \left. \exp \left\lbrace \frac{a y_2}{D}\right\rbrace \frac{\exp \left\lbrace -\frac{(y_2+at)^2}{4Dt}\right\rbrace}{y_2+at} -\exp \left\lbrace \frac{a y_1}{D}\right\rbrace \frac{\exp \left\lbrace -\frac{(y_1+at)^2}{4Dt}\right\rbrace}{y_1+at} \right]
\eeq
We shall consider the case where $y_2 = y_1(1+\eps)$ in (\ref{Sp_for_[y_1, y_2]_with_a_drift}), and  $\eps$ tends to zero, thus
\beq\label{terminos}
S(t) &\sim& S_0(t)+ S_1(t) \eps +  S_2(t)\frac{\eps^2}{2}+...
\eeq
where
\beq
S_0(t) &=& 1-\frac{\sqrt{Dt}\exp\left\lbrace-\frac{y_1^2}{(4Dt)}\right\rbrace}{y_1\sqrt{\pi}}\left(1+e^{\frac{ay_1}{4Dt}}\right)\nonumber \\
S_1(t) &=& \frac{3}{\sqrt{4D\pi t}}\exp\left\lbrace-\frac{(y_1-a t)^2}{(4Dt)}\right\rbrace - \frac{a}{2D}e^{\frac{ay_1}{4Dt}}\mathrm{erfc}\left(\frac{y_1+at}{\sqrt{4Dt}}\right) \nonumber \\
&+&\frac{D(y_1-at)}{2aDt\sqrt{4Dt}}\exp\left\lbrace-\frac{(y_1-a t)^2}{(4Dt)}\right\rbrace \left(1-e^{\frac{ay_1}{4Dt}}\right)\nonumber
\eeq
When $ \eps \rightarrow 0 $ (that is $y_2 \rightarrow y_1 $), we recover in equation (\ref{terminos}), the survival probability of the Dirac delta function $\delta(x-y_1)$, but we cannot get the MFAT due to the exponentially small terms that appear canceling with each other when $\eps$ and $\frac{\eps}{t}$ are small. Thus, to leading order, we have
\beq \label{mfat for y1_y2 with drift}
\overline{\tau}^{N} \sim
\frac{y_1^2}{4D \log\left( \frac{N\left( 1 + \exp \left\lbrace\frac{ay}{D}\right\rbrace \right)}{ 2\sqrt{\pi}}\right)}.
\eeq
}
\section{Discussion and concluding remarks}
In summary, we have obtained several asymptotic formulas for the mean time of the fastest Brownian particles to reach a target. These formulas crucially depend on the initial distribution toward the isolate target: we found algebraic vs $1/\log N$ decay depending on the different initial density profile. In the context of extreme value statistics, the distribution of the first arrival time $\tau^ 1{(N)} $ is the minimum among the random variables
\beq
\tau^1(N) = \min\left(t_1, t_2, \cdots, t_N\right)\geq 0,
\eeq
and thus the limiting distribution of $\tau^1(N)$ is given by the Weibull distribution (because of the lower bound of the support which is positive). This is the case when the target site located at the origin is included in the support of the initial distribution of the particles. Indeed, in that case, the pdf of the $t_i$ behaves as a power law near zero. The Weibull expression implies the algebraic dependence of $\tau^1(N)$ with $N$. However, if the origin is not part of the initial density profile, the pdf of the $t_i$'s has an essential singularity at small argument (see above equation \ref{mfat unif [y_1,y_2] no drif}). Consequently, the limiting form is not a Weibull distribution but instead a Gumbel law, which in turn implies a decay in the form of $\frac{1}{\log N}$, as discussed in the review \cite{majumdar2020extreme} (below equation (25) therein). \\
In general, the renewal interest of extreme statistics \cite{weiss1983order,majumdar2016exact,schehr2014exact} is due to recent direct applications in cell biology \cite{basnayake2019fastestphys,sokolov2019extreme,rusakov2019extreme,coombs2019first} which appears as a frame to explain fast molecular activation. The frame of extreme statistics can be used to compute how the molecular activation time depends on the main parameters, involving the geometrical organization and the dynamics (diffusion or other stochastic processes). The fastest molecules to activate a target site uses the shortest path, thus showing that the redundancy property can overcome the hindrance of a crowded environment.\\
This redundancy principle is ubiquitous in cell biology: One class of example is calcium signalling that can be amplified by activating the calcium-induced-calcium released pathway \cite{fain2019sensory}. This amplification does not require the transport of all ions but only the first ones to arrive to a specific targets made of few clustered receptors. The amplification occurs in few milliseconds instead of hundreds of milliseconds as would be predicted by the time scale of the classical diffusion and the narrow escape theory \cite{Holcman2015} at synapse of neuronal cells \cite{korkotian2017role,korkotian2014synaptopodin,korkotian1999release,basnayake2019fastplos}. Interestingly the arrival time of the fastest among $N$ decays with $\frac{\delta^2}{\log N}$, when the source and the target are well separated by a distance $\delta$. However, there are several situations where choosing the Dirac delta function might not be the best model, as we discussed here. For example, when the particle injection could take a certain time, an extended initial distribution can build up, that could be approximated by a Gaussian or any other related distribution with an algebraic decay, especially when the motion can be modeled as anomalous diffusion (see relation (\ref{genaral_alpha_formula})).\\
Another transduction applications concerns the activation of secondary messengers such as IP$_3^+$ receptors involved in the genesis of calcium wave in astrocytes \cite{rouach2000activity} or the fast activation of TRP channels in fly photo-receptor, which are located very close to the source of the photoconversion.\\
In addition, we obtained here a novel formula when the dynamics contains a local constant flow added to the Brownian component. A local flow could accelerate the transport of the fastest molecules, which could be the case for the delivery occurring inside the endoplasmic reticulum network \cite{dora2018active}. This network is indeed composed of thin tubules well approximated as dimension one segment intersecting at nodes. \\
Finally, it would be interesting to extend the present analysis to the case of exiting from a basin of attraction and study the mean arrival time of the fastest. The case of an OU process is already delicate as there is no exact closed formula for the survival probability with a zero absorbing boundary condition at a given threshold.  Indeed  for an OU process $dx = -\theta x dt +\sqrt{2D}dw$ centered at the origin with an absorbing boundary at $x = 0$ and initial point at $x = y$ and $\theta \geq 0$, the arrival time for the fastest is given at leading order by the same formula as if there was only diffusion and no drift. Indeed in this very particular case, the solution has the form
\beq
p(x,t) = \sqrt{\frac{\theta}{2 \pi D (1- e^{-2 \theta t})}}\left[ \exp \left \lbrace -\frac{\theta}{2D} \frac{(x-ye^{-\theta t})^2}{(1- e^{-2 \theta t})} \right \rbrace - \exp \left \lbrace -\frac{\theta}{2D} \frac{(x+ye^{-\theta t})^2}{(1- e^{-2 \theta t})}\right \rbrace \right]
\eeq
and the survival probability is
\beq
Pr\left \lbrace t_1 >t \right \rbrace  = 1 - \mathrm{erfc} \left(\frac{\sqrt{\theta} y e^{-\theta t}}{ \sqrt{2 D (1- e^{-2 \theta t})}} \right).
\eeq
Then, for $t$ small, we have
\beq
\overline{\tau}^N \sim \frac{y^{2}}{4D \,\,\log\left(\frac{N}{\sqrt{\pi}}\right)}.
\eeq
\\
However for other cases in which the absorbing boundary is not at the maximum point of the parabola, a general approximated formula \cite{martin2019long} has been proposed, correcting erroneous expression found in the literature. At this stage, we could not use their complex formula to estimate the time of the fastest. We speculate that the formula for the mean arrival time for the fastest should be associated not with the Euclidean distance but  with the control problem for the Wencell-Freidlin functional in the Large-Deviation theory, a project for a future investigation.
\section{Appendix: Algorithm to simulate stochastic trajectories of the fastest when the initial distribution can intersect with the target}
{To simulate the arrival of the fastest particle to an absorbing boundary, we use the classical Euler's scheme \cite{schuss2015brownian}. When the source is well separated from the absorbing boundary, we follow each Brownian particle and estimated the time for the first one to arrive. \\
When particles are initially positioned with a distribution that could intersect with the absorbing boundary, the simulation scheme requires more attention, because in principle, particles can be found as close as we wish to the absorbing boundary, and thus the discretization time step could influence the time of the fastest. We thus design the following algorithm:
\begin{enumerate}
\item We generated $N$ initial positions uniformly distributed: $\chi_1,..\chi_N\in [0,y_0]$, where $0$ and $a$ are the absorbing boundaries and $y_0 <a$.
\item The time step $\Delta t$ of the Euler's scheme depends on the shortest distance
\beq
\delta_N=\underset{N}{\min} \left\lbrace |\chi_1|,..,|\chi_N|\right \rbrace,
\eeq
so that the mean square jump is smaller that the shortest distance:
\beq
\Delta t\leq  p\frac{\delta_N^2}{2D},
\eeq
where $D$ is the diffusion coefficient, $p<1$ is a security parameter. In practice, we choose $p=0.2$.
\item For each realization $\omega$, we generated a simulation following step 1 and 2 and computed the first arrival time of the fastest:
\beq
\tau^{N}_{\omega_j} =\inf_{i=1..N} t_{i,j},
 \eeq
where $j = \underset{k}{\inf} \left\lbrace X(k \Delta t) \leq 0 |  X((k-1) \Delta t)>0 \,\,\, \mathrm{or} \,\,\, X(k \Delta t)\geq a |  X((k-1) \Delta t)<a \right \rbrace$.
\item We approximate the mean fastest arrival time by the empirical sums:
\beq
\bar\tau^{N}_m=\frac{1}{m}\sum_{1}^{m}\tau^{N}_{\omega_j}, \text{ with } \bar\tau^N =\underset{m \rightarrow +\infty}{\lim}\bar\tau^{N}_m.
\eeq
\end{enumerate}
}
\normalem
\bibliographystyle{ieeetr}
\bibliography{BiblioCalciumSTIM1-3,references10final,ref_general,RMPbiblio4newN2,biblio,First-passage-timeBibliography,PRLCellsensingbiblio3, biblioPierre}

\end{document}